\documentclass[aps,prl,superscriptaddress,twocolumn,10pt,nofootinbib,preprintnumbers,showdata]{revtex4-2}
\usepackage{bm}
\usepackage{epsfig}
\usepackage{graphics}
\usepackage{amsmath}
\usepackage[usenames, dvipsnames]{color}
\usepackage{colortbl}

\begin{document}

\title{ Self-Consistent Relaxation Theory of Collective Ion Dynamics  \\
in Yukawa One-Component Plasmas \\
under Intermediate Screening Regimes }
\author{ \firstname{Anatolii~V.}~\surname{Mokshin} }
\affiliation{Department of Computational Physics, Institute of Physics, Kazan Federal
University, 420008 Kazan, Russia}
\author{ \firstname{Ilnaz~I.}~\surname{Fairushin}}
\affiliation{Department of Computational Physics, Institute of Physics, Kazan Federal
University, 420008 Kazan, Russia}
\author{ \firstname{Igor~M.}~\surname{Tkachenko}}
\affiliation{Departament de Matem\`{a}tica Aplicada, Universitat Polit\`{e}cnica de Val%
\`{e}ncia, Camino de Vera s/n, 46022 Valencia, Spain}
\affiliation{Al-Farabi Kazakh National University, al-Farabi Av. 71, 050040 Almaty, Kazakhstan}

\begin{abstract}
The self-consistent relaxation theory is employed to describe the collective ion dynamics in
strongly coupled Yukawa classical one-component plasmas. The theory is applied to equilibrium states corresponding to intermediate screening regimes with appropriate values of the structure and coupling parameters. The information about the structure (the radial distribution function and the static structure factor) and the thermodynamics of the system is sufficient to describe collective dynamics over a wide range of spatial ranges, namely, from the extended hydrodynamic to the microscopic dynamics scale. The main experimentally measurable characteristics of the equilibrium collective dynamics of ions -- the spectrum of the dynamic structure factor, the dispersion parameters, the speed of sound and the sound attenuation -- are determined within the framework of the theory without using any adjustable
parameters. The results demonstrate agreement with molecular dynamics simulation results. Thus a direct realization is presented of the key idea of statistical mechanics: for the theoretical description of the collective dynamics of equilibrium fluids it is sufficient to know the interparticle interaction potential and the structural characteristics. Comparison with alternative or complementary theoretical approaches is provided.
\end{abstract}

\maketitle

\affiliation{Department of Computational Physics, Institute of Physics, Kazan Federal
University, 420008 Kazan, Russia}

\affiliation{Department of Computational Physics, Institute of Physics, Kazan Federal
University, 420008 Kazan, Russia}

\affiliation{Departament de Matem\`{a}tica Aplicada, Universitat
Polit\`{e}cnica de Val\`{e}ncia, Camino de Vera s/n, 46022 Valencia, Spain}

Collective dynamics determines essential physical properties of a
many-particle system including the sound propagation, the heat capacity, and
the mass and heat transfer. If we consider a crystalline solid as a system
of interacting particles, as it is customary done in statistical mechanics,
then the concept of phonons is applied to describe the collective dynamics
of such systems~\cite{Landau_Lifshitz_1980}, and qualitative results can be
achieved if the interparticle interaction potential and the structural
characteristics (the lattice type and the lattice constant) are initially
known. However, the extension of the concept of phonons to collective
particle dynamics in equilibrium classical liquids is somewhat misleading or
dubious. Instead of this, in the case of liquids, the time correlation
function formalism appears to be sufficiently efficient~\cite{Fisher_1964}. This
formalism serves as a suitable basis for various theories that are proposed
for reproducing collective and single-particle dynamics in liquids such as
the generalized hydrodynamics~\cite{Hansen/McDonald_book_2006}, the
viscoelastic theory~\cite{Balucani,Boon/Yip_1991}, the mode-coupling theory~%
\cite{Gotze_book} and others. Despite an impressive progress in the
theoretical interpretation of currently available experimental results
related to the collective dynamics in liquids~\cite%
{Price_RPP_2003,Scopigno_RMP_2005,Gaspard_review}, an appropriate theory is
still missing, which could be based solely on an interaction potential and
structural characteristics as input parameters~\cite%
{Tkachenko_PRL_2016,Murillo_2019}.

It is remarkable that a model of many-particle system, where particles interact
through the Yukawa (screened-Coulomb, or Debye-H\"{u}ckel) potential
\begin{equation}
\phi (r)=\frac{Q^{2}}{r}\exp \left( -\frac{r}{\lambda _{s}}\right) ,
\label{eq: Yukawa_potential}
\end{equation}%
is a very suitable for the advancement and testing of such a theory.
Here, $Q$ is the effective particle charge and $\lambda _{s}$ is the
Debye screening length~\cite{Horn_1969}. The interparticle interaction defined by Eq.~(\ref{eq: Yukawa_potential}) reproduces the repulsion of point ions neutralized by the (electronic) background, and such an interaction
corresponds to the case of the classical one-component plasma (OCP) called
the Yukawa-OCP~\cite{Khrapaks_Phys_Plasma_2014,Khrapak_2,Khrapak_3,Ichimaru}. If the
microscopic structure is known (for example, in terms of the pair
distribution function of the particles), then the simple analytical form of
the Yukawa potential allows one to find expressions for the internal and
free energies, the internal pressure, the shear stress, and also to
determine the system virial equation of state~\cite{Hansen/McDonald_book_2006}. Note that the Yukawa-OCP is
of interest not only from the point of view of the fundamental issues of
liquid matter physics, but it has also remarkable applications in various
physical situations, including interiors of neutron stars and white dwarfs,
dusty plasmas, ultra-cold plasmas and colloidal suspensions~\cite%
{Hansen_Lowen,Fortov_reviews_1, Fortov_reviews_2, Killian_reviews,Kremer, Bonitz_PRL_1, Bonitz_PRL_2, Murillo}.

The dynamic structure factor $S(k,\omega )$ ($k$ being the wavenumber, and $%
\omega $ being the frequency) contains a complete information about the
collective particle movements in a many-particle system. This quantity is
experimentally measurable by inelastic scattering of light, neutrons and
X-rays. On the other hand, the dynamic structure factor can be directly
calculated from known particle trajectories, which were initially determined
by experimental methods or by means of molecular dynamics (MD) simulations.
In addition, the quantity $S(k,\omega )$ is the Fourier transform in time of
the density-density correlation function $F(k,t)$ known also as the
intermediate scattering function~\cite{Hansen/McDonald_book_2006}.
Therefore, it is usually possible to compute $S(k,\omega )$ theoretically
and a direct comparison of theoretical and experimental results for the
dynamic structure factor can verify the validity and quality of a theory
suggested to describe the collective dynamics of system particles.

The dynamic properties of the Yukawa-OCP in the intermediate screening regime, i.e. with the finite screening lengths in interparticle interactions, are characterized by the presence of propagating waves, manifested in the dynamic structure factor spectra as a shifted-frequency doublet. This feature is symmetric and typical for dense classical liquids. The dispersion of these collective excitations is similar to that usually observed in equilibrium dense liquids and is completely different from the one characteristic for the Coulomb systems, i.e. when in Eq.~(\ref{eq: Yukawa_potential}) the screening length $\lambda_s$ tends to infinity. A
remarkable feature of the collective dynamics of the Yukawa-OCP is the
practically absent zero-frequency Rayleigh mode in the extended hydrodynamic
wavenumber range. This mode is associated with the non-propagating isobaric
entropy fluctuations~\cite{Landau_Lifshitz_1980}. In the present paper, we
wish to demonstrate that all main characteristics of the classical
Yukawa-liquid collective dynamics for the whole wavenumber range and in the intermediate screening
regime can be determined in a self-consistent manner within the relaxation (microscopic)
theory.

The simplest way to treat an experimental scattering law $S(k,\omega )$ is
to fit this spectrum by a linear combination of some model functions, whose
parameters are identified with some physical parameters. However, physically
justified reasons for such a fit can be given for two limiting cases only:
the low-$k$ (long-range) hydrodynamic and the high-$k$ (short-range) free
particle dynamics limits. There is an alternative to these fittings.

Any scattering law $S(k,\omega )$ at a fixed $k$ can be characterized by a
set of its frequency moments
\begin{equation}
\langle \omega ^{(l)}(k)\rangle =\frac{1}{\rho }\int_{-\infty }^{\infty
}\omega ^{l}\,S(k,\omega )\,d\omega ,\ \ \ l=0,\;1,\;2,\;\ldots ,
\label{eq: freq_parameters}
\end{equation}%
usually called the sum rules, $\rho $ is the particle concentration. The
dimension of the $l$-th order frequency moment depends on its order $l$, and
it is (\textit{frequency})$^{l}$. Therefore, it is more convenient to use
the set of frequency parameters defined by the ratios of the moments:
\begin{eqnarray} \label{eq: relaxation_parameters}
\Delta _{1}(k) &=&\frac{\langle \omega ^{(2)}(k)\rangle }{\langle \omega
^{(0)}(k)\rangle }, \\
\Delta _{2}(k) &=&\frac{\langle \omega ^{(4)}(k)\rangle }{\langle \omega
^{(2)}(k)\rangle }-\frac{\langle \omega ^{(2)}(k)\rangle }{\langle \omega
^{(0)}(k)\rangle },  \notag \\
\Delta _{3}(k) &=&\frac{\left[ \langle \omega ^{(6)}(k)\rangle \langle
\omega ^{(2)}(k)\rangle -\left( \langle \omega ^{(4)}(k)\rangle \right) ^{2}%
\right] \langle \omega ^{(0)}(k)\rangle }{\langle \omega ^{(4)}(k)\rangle
\langle \omega ^{(2)}(k)\rangle \langle \omega ^{(0)}(k)\rangle -\left(
\langle \omega ^{(2)}(k)\rangle \right) ^{3}},  \notag
\end{eqnarray}%
\begin{equation*}
\ldots ,
\end{equation*}%
having all the same dimension of the frequency squared. In a classical
system the dynamic structure factor is an even function of frequency so that
its odd-order moments vanish and the $n$-th order frequency parameter $%
\Delta _{n}(k)$, where $n=1$, $2$, $3$, $\ldots$, can be expressed in terms of only even-order moments up to a
$2n$-th order one: 
$\langle \omega^{(2)}(k)\rangle $, $\langle \omega^{(4)}(k)\rangle $, $\ldots $, $\langle
\omega^{(2n)}(k)\rangle $. It is remarkable that the frequency parameters
can be determined independently and exactly in terms of the microscopic
characteristics. In particular, for the first frequency parameter due to the
fluctuation-dissipation theorem one has that
\begin{subequations}
\label{eq: freq_parameters_microsc}
\begin{equation}
\Delta _{1}(k)=\frac{\omega _{p}^{2}}{3\Gamma }\frac{(ka)^{2}}{S(k)},
\label{eq: Delta_1}
\end{equation}%
while the second frequency parameter can be written as
\begin{equation}
\Delta _{2}(k)=\Delta _{1}(k)\left[ 3S(k)-1\right] +\omega _{p}^{2}\,D(k)
\label{eq: Delta_2}
\end{equation}%
with (some details are provided in the Supplemental Material~\cite{Suppl})
\end{subequations}
\begin{eqnarray}
D(k) &=&\int_{0}^{\infty }\frac{\exp (-\kappa x)}{3x}\bigl[\left( 2(\kappa
x)^{2}+6\kappa x+6\right) j_{2}(kax)  \notag \\
&&+(\kappa x)^{2}(1-j_{0}(kax))\bigr]g(x)dx,  \notag
\end{eqnarray}%
where
\[
    S(k) \equiv \langle \omega ^{(0)}(k)\rangle = 1 + \frac{4 \pi \rho}{k} \int_{0}^{\infty} r \left [ g(r) -1 \right ] \sin(kr)dr
\]
\noindent
is the static structure
factor, $\omega _{p}=~\sqrt{4\pi Q^{2}\rho /m}$ is the plasma frequency, $a=
\sqrt[3]{3/(4\pi \rho )}$ is the Wigner-Seitz radius, $\Gamma
=Q^{2}/(k_{B}Ta)$ is the coupling parameter, $\kappa =~a/\lambda _{s}$ is
the structure parameter, $x=r/a$ is the dimensionless spacial variable, $%
j_{n}(x)$ are the spherical Bessel functions, $g(x)$ is the radial
distribution function. Generally speaking, the expression for the $n$-th order frequency parameter $\Delta_n(k)$ at $n \geq 2$ contains an integral expression with the interparticle potential and the distribution function for $n$ particles.

The higher the order of the frequency moment (or the
relaxation parameter), the more high-frequency properties of the spectrum it
captures. The set of these moments (and the relaxation parameters) uniquely
matches a specific spectrum at a fixed $k$. Therefore, it is reasonable to
expect that the scattering law can be expressed in terms of the relaxation
parameters as a functional. Based on the set of the dynamic variables
\begin{equation}
\mathbf{W}(k)=\{W_{0}(k),\ W_{1}(k),\ W_{2}(k),\ W_{3}(k),\ W_{4}(k)\}
\end{equation}%
interrelated by the following recurrent relations
\begin{eqnarray}
W_{j+1}(k) &=&\frac{d{W}_{j}(k)}{dt}+\Delta _{j}(k)W_{j-1}(k),  \notag \\
\Delta _{j}(k) &=&\frac{\langle |W_{j}(k)|^{2}\rangle }{\langle
|W_{j-1}(k)|^{2}\rangle },\ W_{-1}(k)\equiv 0,  \notag \\
&& j=0,\,1,\,2,\,\ldots, \notag
\end{eqnarray}%
where the initial variable $W_{0}(k)=\rho _{k}$ defines the local density
fluctuations, the self-consistent relaxation theory provides the dynamic
structure factor in terms of the first four relaxation parameters, $%
S(k,\omega )\propto \mathcal{F}\left[ S(k),\,\Delta _{1}(k),\,\Delta
_{2}(k),\,\Delta _{3}(k),\,\Delta _{4}(k)\right] $, where $\mathcal{F}\left[
\ldots \right] $ means an algebraic expression~\cite{AVM_PRE_2001,AVM_JPCM_2003}.
The exact equation for the dynamic structure factor $S(k,\omega)$, obtained on the basis of the self-consistent relaxation theory, can be found in Refs.~\cite{Mokshin_TMF_2015,Mokshin/Galimzyanov_JPCM_2018,Mokshin_JCP_2004} (see, for example, Eq.~(42) in Ref.~\cite{Mokshin/Galimzyanov_JPCM_2018}).
We notice that the relaxation theory belongs to the theoretical schemes, where the known infinite chain of integrodifferential equations for the time correlation functions of variables from the set $\mathbf{W}(k)$ is solved in a self-consistent way as, for example, in the self-consistent mode-coupling theory~\cite{Gotze_1984,Miyazaki_2004,Takeno_1979,PRE_2019,Reichman_2005}, and no approximations of these time correlation functions by any model functions with free parameters are required~\cite{Szamel}. This also becomes possible when the entire infinite set of sum rules (\ref{eq: freq_parameters}) [or (\ref{eq: relaxation_parameters})] is known~\cite{Mokshin/Galimzyanov_JPCM_2018,Mokshin_TMF_2015,AVM_TMPh_2021,PRB_2020,Florencio_RRM}. In the case we consider here, the theoretical procedure is non-perturbative, which is especially appropriate for the description of the systems we are dealing with here. The main ideas of the theory are to take advantage of the correspondence between the time-scales of a sequence of relaxation processes associated with the dynamical variables from the set $\mathbf{W}(k)$, and of the fact that the time-scales themselves are evaluated through the frequency parameters as $\tau_n(k)=\Delta_n^{-1/2}(k)$ \cite{AVM_PRL_2005}. The description is based solely on the assumption that relaxation processes, determined by the energy flow, and by more subtle physical effects that are determined through the derivatives of the energy current with respect to time, occur on higher-order time-scales $\tau_n(k)$ which become asymptotically equal at large $n$. In the case of classical equilibrium fluids independently of the interaction range of the fluid particles, this condition is sufficient to find the dynamic structure factor as well as other characteristics of the collective dynamics of particles. However, for specific systems, one can expect to find additional interrelations between frequency parameters, which will cause corresponding modifications in the relaxation theory.

\begin{figure}[h]
\centering
\includegraphics[keepaspectratio,width=\linewidth]{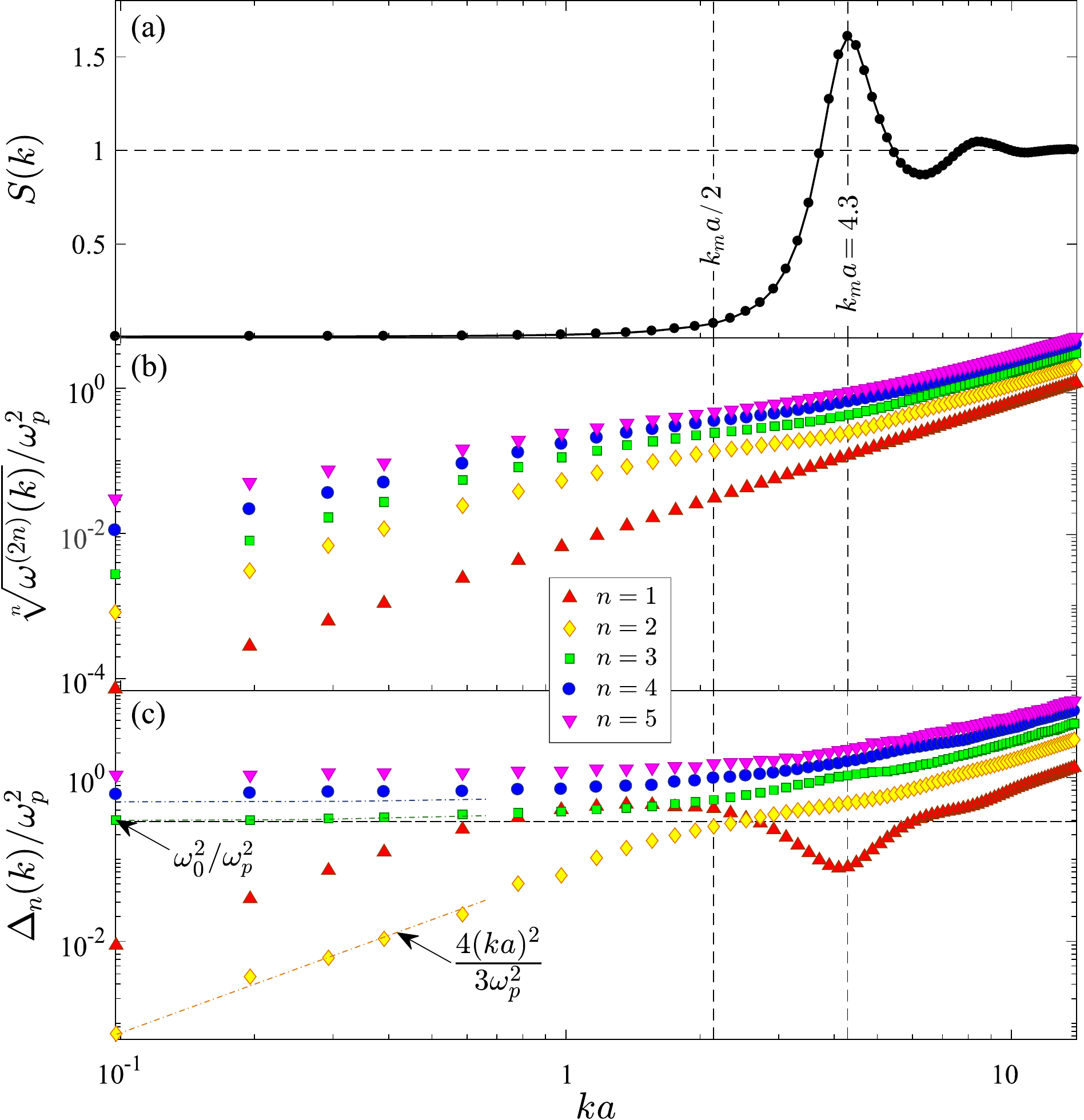} 
\caption{Wavenumber dependence of some characteristics evaluated by means
of molecular dynamics simulations for the Yukawa-OCP at the thermodynamic
state with $\Gamma = 50$ and $\protect\kappa = 1$: (a) dots connected by the solid line represent the static structure
factor $S(k)$, where the location of the main peak is marked by $k_m a$; and
$k_m a/2$ corresponds to the first pseudo-Brillouin zone boundary; (b)
reduced frequency moments; (c) frequency parameters scaled by $\protect\omega%
_p^2$. Low-$k$ asymptotic forms shown by straight lines, for $\Delta_1(k)$ it follows
from Eq.~(\protect\ref{eq: Delta_1}), for $\Delta_2(k)$ it can be obtained
taking into account the results of the quasi-localized charge
approximation~(QLCA)~\protect\cite{Kalman, Golden, Klumov, Khrapak2019,
Fairushin}, and for the parameters $\Delta_3(k)$ and $\Delta_4(k)$ they are obtained
empirically.}
\label{Fig_Moments_new}
\end{figure}
\begin{figure}[h]
\centering
\includegraphics[keepaspectratio,width=\linewidth]{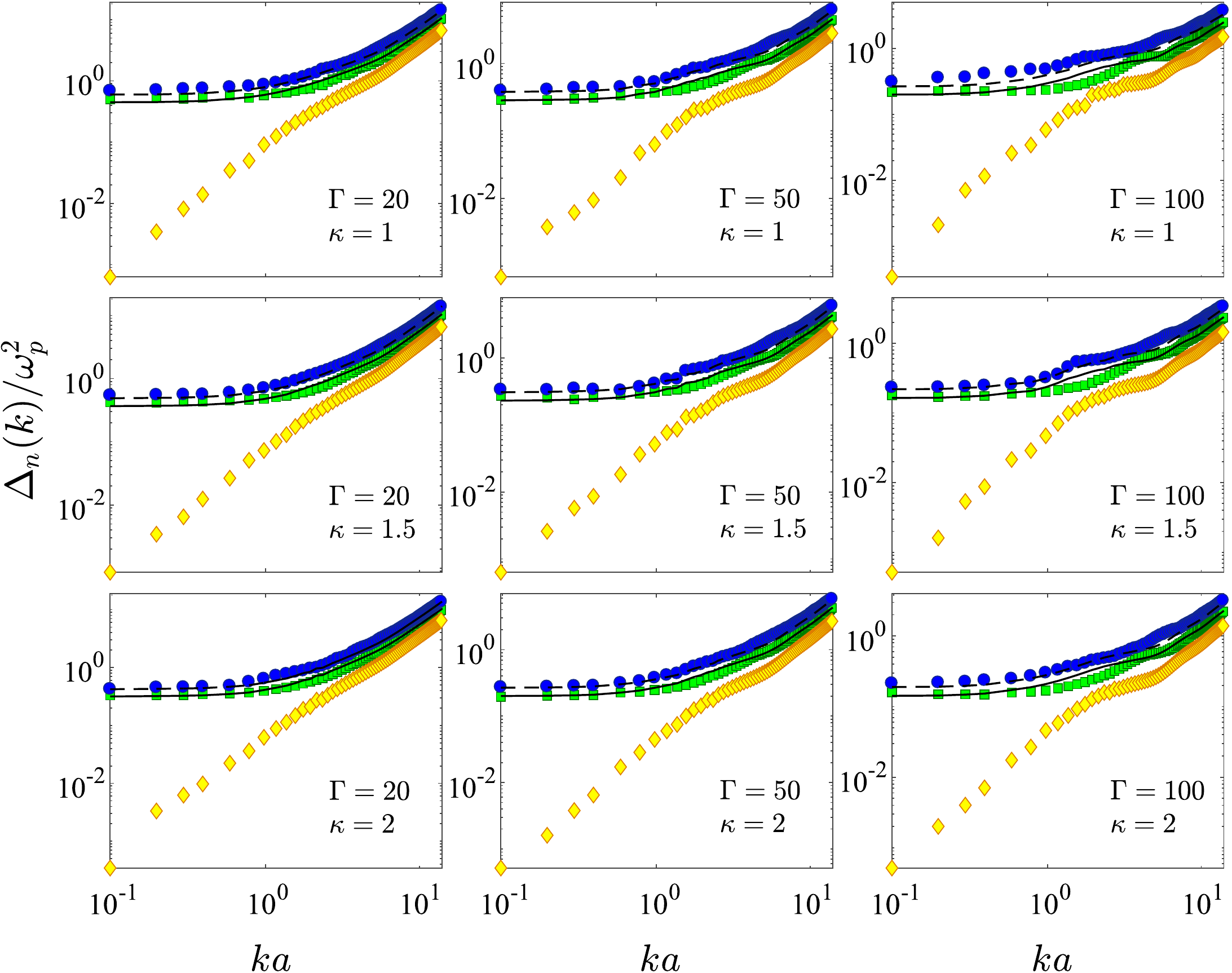}
\caption{Wavenumber dependence of the scaled frequency parameters $\Delta_2(k)/\omega_p^2$ (yellow diamonds), $\Delta_3(k)/\omega_p^2$ (green squares) and $\Delta_4(k)/\omega_p^2$ (dark blue circles) obtained by molecular dynamics simulations of the Yukawa-OCP for nine thermodynamic states and some combinations of $\Gamma$ and $\kappa$. Solid and dashed lines correspond to relations~(\ref{eq: Delta_3}) and (\ref{eq: Delta_4}), respectively.
}
\label{fig: Delta3_4}
\end{figure}
\begin{figure}[t]
\centering
\includegraphics[keepaspectratio,width=1.2\linewidth]{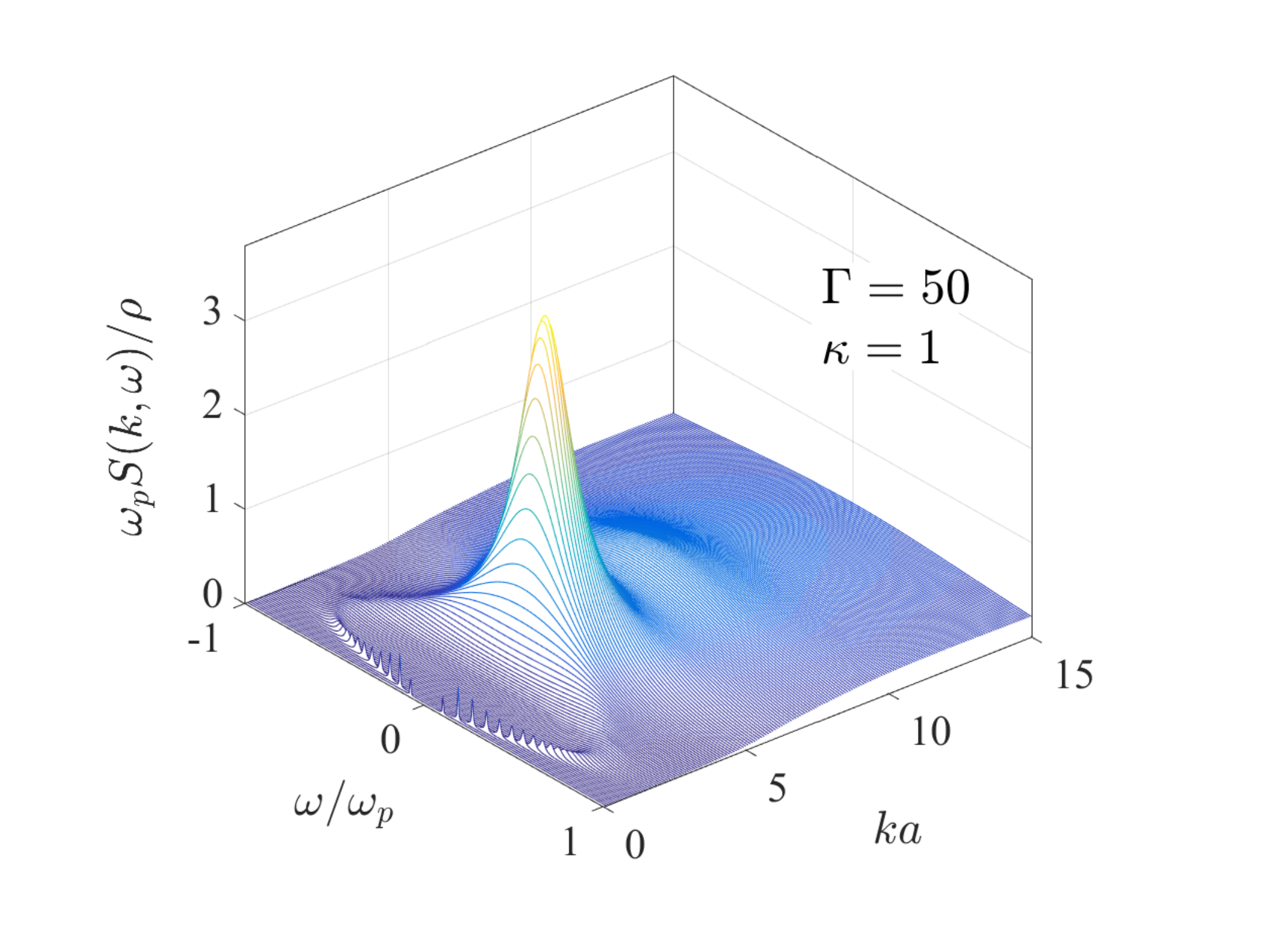} 
\caption{Dynamic structure factor vs. the wavenumber $ka$ and the frequency $\omega/\omega_p$ at the thermodynamic state with $\Gamma=50$ and $\kappa=1$ evaluated from Eq.~(\ref{eq: S_k_omega}). Note that the radial distribution function $g(r)$ generated on the basis of independent molecular dynamics simulations is used as a sole input parameter to compute the dynamic structure factor via Eq.~(\ref{eq: S_k_omega}).}
\label{fig: dsf}
\end{figure}

The equilibrium molecular dynamics simulations~\footnote{The equilibrium molecular dynamics simulations of the Yukawa-OCP for $\Gamma=20$, $50$, $100$ and $\kappa=1$, $1.5$, $2$ were carried out using the computational package LAMMPS [S. Plimpton, \textsl{J. Comput. Phys.} \textbf{117}, 1 (1995)]. The simulation cell contained $64\,000$ particles interacting with the Yukawa potential, and periodic boundary conditions were applied to the cell in all directions. The evolution of the system corresponding to the $NVT$ ensemble was monitored. The particle motion equations were integrated in accordance with the Verlet algorithm with a time integration step $\tau =0.01/\omega_p$.} of the Yukawa-OCP for $\Gamma
=20,\,50,\,100$ and $\kappa =1,\,1.5,\,2$ reveal no simple correlation
between the frequency parameters $\Delta _{1}(k)$ and $\Delta _{2}(k)$, and,
therefore, it is not possible to simplify Eq.~(\ref{eq: Delta_2}).
Nevertheless, there is a correspondence between $\Delta _{2}(k)$, $\Delta
_{3}(k)$ and $\Delta _{4}(k)$ for the extended range of the wavenumber
variation, that is clearly seen from the defined $k$-dependence of these
frequency parameters (Figs.~\ref{Fig_Moments_new} and \ref{fig: Delta3_4}):
\begin{subequations}
\label{eq: Delta_s}
\begin{equation}   \label{eq: Delta_3}
    \Delta _{3}(k)=\frac{3}{2}\Delta _{2}(k)+\omega _{0}^{2},
\end{equation}
\begin{equation} \label{eq: Delta_4}
    \Delta _{4}(k)=\frac{4}{3}\Delta _{3}(k)=2\Delta _{2}(k)+\frac{4}{3}%
    \omega _{0}^{2}
\end{equation}
\end{subequations}
with
\begin{equation*}
\omega _{0}^{2}=\frac{2\omega _{p}^{2}}{\sqrt{\Gamma \kappa }}.
\end{equation*}
Relations~(\ref{eq: Delta_s}) express the higher-order relaxation parameters $\Delta_3(k)$ and $\Delta_4(k)$ in terms of the parameter $\Delta_2(k)$. These relations are similar, in a sense, to the representation of the three- and four-particle distribution functions in terms of the pair correlation function -- the radial distribution function of particles.
Further, relations~(\ref{eq: Delta_s}) satisfy the Cauchy-Bunyakovsky-Schwarz
inequalities for the frequency moments~\cite{TkachenkoPRE2014,
TkachenkoPRE2020}:
\begin{eqnarray}
\frac{\langle \omega ^{(2)}(k)\rangle }{\langle \omega ^{(0)}(k)\rangle }
&\leq &\frac{\langle \omega ^{(4)}(k)\rangle }{\langle \omega
^{(2)}(k)\rangle },\ \ \frac{\langle \omega ^{(4)}(k)\rangle }{\langle
\omega ^{(0)}(k)\rangle }\leq \frac{\langle \omega ^{(8)}(k)\rangle }{%
\langle \omega ^{(4)}(k)\rangle },  \label{eq: Cauchy} \\
\frac{\langle \omega ^{(6)}(k)\rangle }{\langle \omega ^{(4)}(k)\rangle }
&\leq &\frac{\langle \omega ^{(8)}(k)\rangle }{\langle \omega
^{(6)}(k)\rangle }.  \notag
\end{eqnarray}
Notice that these inequalities warrant the correct mathematical structure and properties of our results, e.g. the positiveness of the dynamic structure factor and of the decrement of the collective modes, etc. Besides, their fulfillment implies the compliance of the present approach with the fundamental requirements.
Taking relations~(\ref{eq: Delta_s}) into account, the self-consistent
relaxation theory yields the dynamic structure factor in the form
\begin{equation}
S(k,\omega )=\frac{\rho S(k)}{\pi }\frac{2\Delta _{2}(k)\sqrt{A_{3}(k)}}{%
\omega ^{6}+A_{1}(k)\omega ^{4}+A_{2}(k)\omega ^{2}+A_{3}(k)},
\label{eq: S_k_omega}
\end{equation}
where
\begin{eqnarray}
A_{1}(k) &=&3\omega _{0}^{2}-\frac{\Delta _{2}(k)}{2}-2\Delta _{1}(k),
\notag \\
A_{2}(k) &=&\left[ \Delta _{1}(k)-2\Delta _{2}(k)\right] ^{2}-6\Delta
_{1}(k)\omega _{0}^{2},  \notag \\
A_{3}(k) &=&\frac{3}{2}\Delta _{1}^{2}(k)\left [ 3\Delta _{2}(k)+2\omega
_{0}^{2}\right ] .  \notag
\end{eqnarray}

\begin{figure*}[t]
\begin{center}
\includegraphics[keepaspectratio,width=\linewidth]{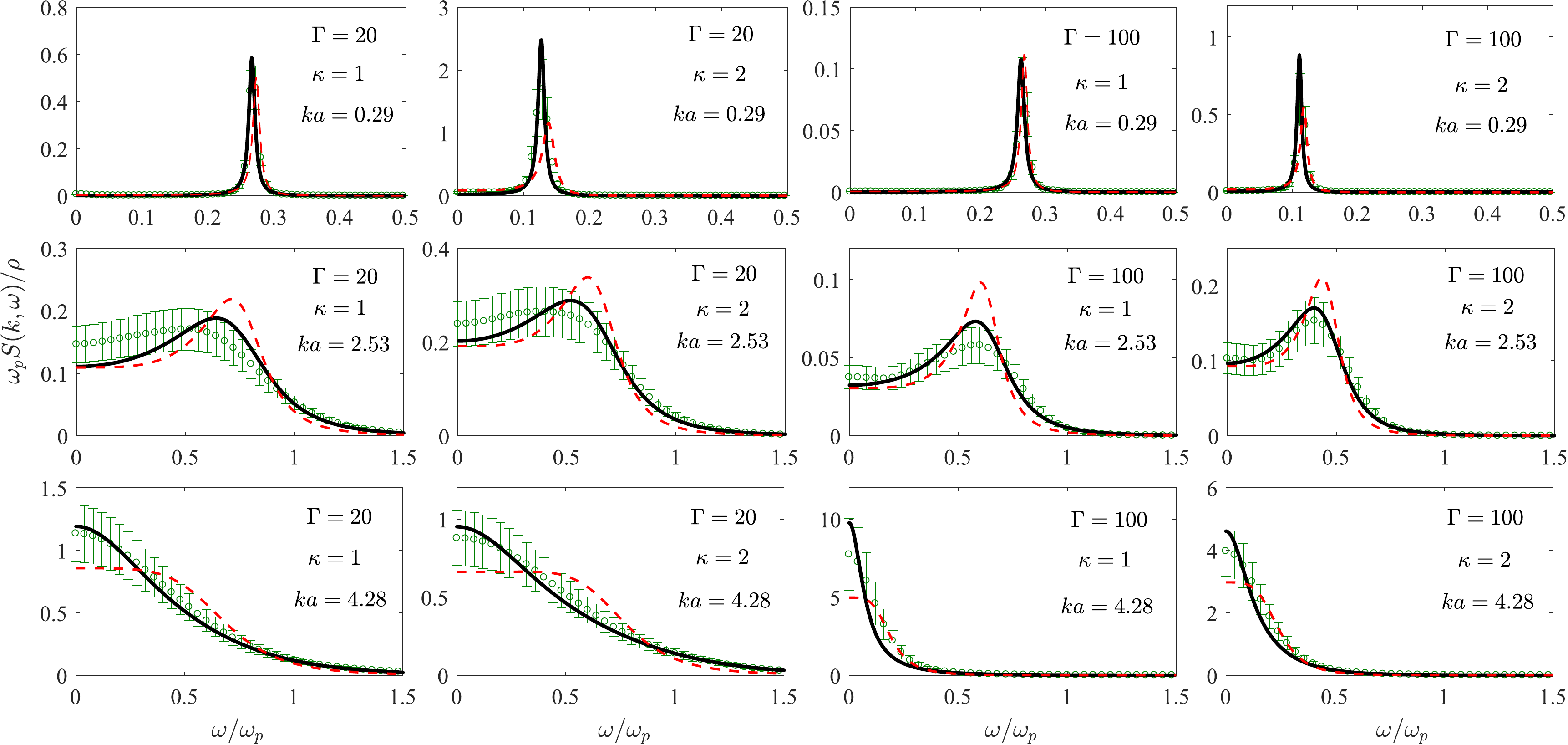}
\end{center}
\caption{Dynamic structure factor spectra, multiplied by the plasma
frequency, for the thermodynamic states at various values of $\Gamma $ and $%
\protect\kappa $ and at the various wavenumbers. Here, theoretical results
from Eq.~(\protect\ref{eq: S_k_omega}) shown by black solid lines are compared
with molecular dynamics (MD) simulation data given by green circles and with
results of the FM-theory~\protect\cite{Tkachenko_PRL_2016} presented by red
dashes lines.}
\label{Skw_1}
\end{figure*}
Some remarkable points associated with relations~(\ref{eq: Delta_s}) and
Eq.~(\ref{eq: S_k_omega}) are to be pointed out. First, relations~(\ref{eq:
Delta_s}) can provide a correct result for the high-$k$ free particle
dynamics limit with $\Delta _{2}(k)\gg \omega _{0}^{2}$ and the following
recurrence relation:
\begin{equation}
\Delta _{n+1}(k)=\frac{n+1}{n}\Delta _{n}(k),\ \ n=1,\;2,\;3,\;\ldots ,
\label{eq: Delta_free_particle}
\end{equation}%
which exactly corresponds to the dynamic structure factor of the Gaussian
form
\begin{equation}
S_{\mathrm{fmp}}(k,\omega )=\sqrt{\frac{\rho ^{2}}{2\pi \Delta _{1}(k)}}\exp
\left( -\frac{\omega ^{2}}{2\Delta _{1}(k)}\right) .
\label{eq: S_k_omega_free_part}
\end{equation}%
Note that Eq.~(\ref{eq: S_k_omega_free_part}) reproduces the dynamic
structure factor spectrum for the regime of \textquotedblleft a free-moving
particle\textquotedblright ~\cite{Hansen/McDonald_book_2006}. Second,
according to Eq.~(\ref{eq: S_k_omega}), the shape of $S(k,\omega )$ at a
fixed $k$ is determined by the bicubic polynomial in the variable $\omega $.
Analysis of~(\ref{eq: S_k_omega}) allows one to obtain the dispersion
equation for the high-frequency quasi-acoustic mode
\begin{eqnarray}
s^{3}+B(k)s^{2} &+&\left[ \Delta _{1}(k)+\frac{8}{5}\Delta _{2}(k)\right]
s+B(k)\Delta _{1}(k)=0,  \notag \\
B(k) &=&\frac{4\sqrt{A_{3}(k)}}{5\Delta _{1}(k)}.
\end{eqnarray}%
Solution of this equation yields $s(k)=\pm i\omega _{c}(k)-\delta (k)$ with
the dispersion for the high-frequency peak of the dynamic structure factor
\begin{equation}
\omega _{c}(k)=\sqrt{3}\left( \sqrt[3]{Z(k)-q(k)}+\sqrt[3]{Z(k)+q(k)}\right)
,  \label{eq: omega_c}
\end{equation}%
the low-$k$ asymptotes of this dispersion
\begin{eqnarray}
v_{s}\,k &=&\lim_{k\rightarrow 0}\omega _{c}(k) \\
&=&\lim_{k\rightarrow 0}\sqrt{\frac{6\left[ (\Delta _{1}(k)+2\Delta
_{2}(k))^{2}-\Delta _{1}(k)\omega _{0}^{2}\right] }{4\Delta _{1}(k)+\Delta
_{2}(k)-6\omega _{0}^{2}}}  \notag
\end{eqnarray}%
and the dispersion for the sound decrement
\begin{equation}
\delta (k)=\sqrt[3]{Z(k)+q(k)}-\sqrt[3]{Z(k)-q(k)}-\frac{B(k)}{3},
\label{eq: sound_att}
\end{equation}%
where
\begin{eqnarray}
Z(k) &=&\sqrt{p^{3}(k)+q^{2}(k)},  \notag \\
p(k) &=&\frac{\Delta _{1}(k)}{12}+\frac{4}{75}\left[ \Delta _{2}(k)-\omega
_{0}^{2}\right] ,  \notag \\
q(k) &=&\frac{B(k)}{600}\left[ 25\Delta _{1}(k)-12\Delta _{2}(k)+\frac{%
16\omega _{0}^{2}}{3}\right] ,  \notag
\end{eqnarray}%
and $v_{s}$ is the sound velocity.

The spectral density $C_{L}(k,\omega )$ for the longitudinal current
correlation function is also determined by the dynamic structure factor $%
S(k,\omega )$:
\begin{equation}
C_{L}(k,\omega )=\frac{\omega ^{2}}{S(k)\Delta _{1}(k)}S(k,\omega ).
\label{eq: CL}
\end{equation}%
Then, taking into account relation~(\ref{eq: S_k_omega}), one can obtain
directly the analytical expression for the spectral density $C_{L}(k,\omega
) $ and find the dispersion relation $\omega _{L}(k)$ for the longitudinal
acoustic-like excitations:
\begin{equation}
\omega _{L}(k)=\sqrt{C_{+}(k)+C_{-}(k)-\frac{A_{1}(k)}{6}}
\label{eq: omega_L}
\end{equation}%
with
\begin{equation}
C_{\pm }(k)=\sqrt[3]{\frac{A_{3}(k)}{4}-\frac{A_{1}^{3}(k)}{216}\pm \sqrt{%
\frac{A_{3}^{2}(k)}{16}-\frac{A_{3}(k)A_{1}^{3}(k)}{432}}}.  \notag
\end{equation}%
Since the parameters $\Delta _{1}(k)$ and $\Delta _{2}(k)$ can be calculated
analytically by means of Eqs.~(\ref{eq: Delta_1}) and (\ref{eq: Delta_2}),
respectively, fitting is not necessary to compute the dynamic structure
factor $S(k,\omega )$ and all other characteristics of the collective
particle dynamics.

Now we check to which extend the theoretical formalism is consistent with
the molecular dynamics simulation data and results of alternative
theoretical approaches. Several theoretical models are known that have been
suggested earlier to describe the collective dynamics of Yukawa classical
one-component plasmas~\cite{Ortner,Adamyan,HansenOCP,Mithen_PRE_2011,Tkachenko_PRL_2016,TkachenkoPRE2020, TkachenkoPRE2014, MithenPRE2012}.
Accurate theoretical description has been provided earlier by the theory based on the method of frequency moments ‒- the FM-theory (for details, see Refs.~\cite{Tkachenko_PRL_2016,TkachenkoPRE2020}). This FM-theory yields the expression for the dynamic structure factor $S(k,\omega )$ in the form of a linear-fractional transformation of the Nevanlinna parameter function (NPF) possessing specific mathematical properties, which guarantee the satisfaction of an imposed set of sum rules or power frequency moments automatically and independently of the NPF model. In Refs.~\cite{Tkachenko_PRL_2016,TkachenkoPRE2020}, the NPF determined by the relaxation frequency parameters $\Delta_1(k)$ and $\Delta_2(k)$  was found  on the basis of physical considerations, and it leads to an expression for $S(k,\omega )$ similar to Eq.~(\ref{eq: S_k_omega}). Remarkably, at certain conditions, Eq.~(\ref{eq: S_k_omega}) transforms into the dynamic structure factor $S(k,\omega)$ of FM-theory given in the above papers exactly.
Details of the interrelation between the present relaxation and the moment self-consistent theoretical approaches is provided in the Supplemental Material~\cite{Suppl}.

For the thermodynamic state with $\Gamma =50$ and $\kappa =1$, the first
maximum of the static structure factor $S(k)$ is located at the wavenumber $k_{m}=4.3\;a^{-1}$ [see Fig.~\ref{Fig_Moments_new} (a)]. The basic features
of the microscopic collective dynamics appear at the wavenumbers $k\in
(0;\;k_{m})$. This can be seen from Fig.~\ref{fig: dsf}, which presents the results predicted by Eq.~(\ref{eq: S_k_omega}) for this thermodynamic state -- the scaled dynamic structure
factor $\omega _{p}S(k,\omega )/\rho$ as function of the scaled wavenumber $ka$ and frequency $\omega/\omega_p$.
The Brillouin doublet in $\omega _{p}S(k,\omega )/\rho$ is seen as symmetric maxima located at non-zero frequencies for the wavenumbers up to $k \simeq 4.0\, a^{-1}$.

In Fig.~\ref{Skw_1}, we show the scaled dynamic structure
factor $\omega _{p}S(k,\omega )/\rho$ computed within the self-consistent
relaxation theory with Eq.~(\ref{eq: S_k_omega}) for the fixed scaled wave
numbers $ka=0.29$, $2.53$ and $4.28$ at the thermodynamic conditions of the
Yukawa-OCP with $\Gamma =20$ and $100$, and $\kappa =1$, $1.5$ and $2$. For
these thermodynamic conditions and wavenumbers, the self-consistent
relaxation theory reproduces the MD simulation results quite accurately and
describes all the spectral features. At small wavenumbers $k\;<k_{m}/2$
corresponding to an extended hydrodynamic range, the spectra of $S(k,\omega
) $ contain just a high-frequency Brillouin component. With the wavenumber $k$ increase\ starting from the values comparable with $k_{m}/2$, the
zero-frequency Rayleigh component emerges and becomes pronounced, while the
high-frequency Brillouin component disappears. As it is seen, Eq.~(\ref{eq:
S_k_omega}) provides sometimes even better agreement with the MD simulation
results than the FM-theory.

\begin{figure}[h]
\centering
\includegraphics[keepaspectratio,width=\linewidth]{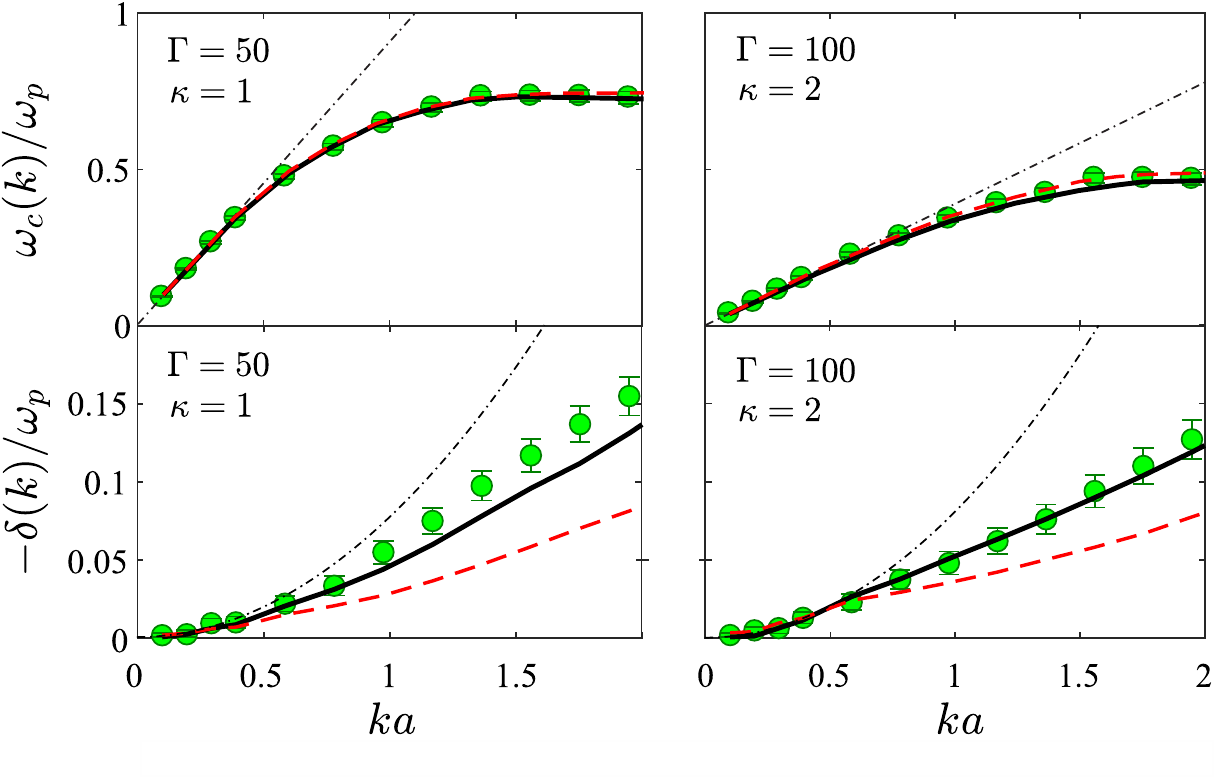} %
\includegraphics[keepaspectratio,width=\linewidth]{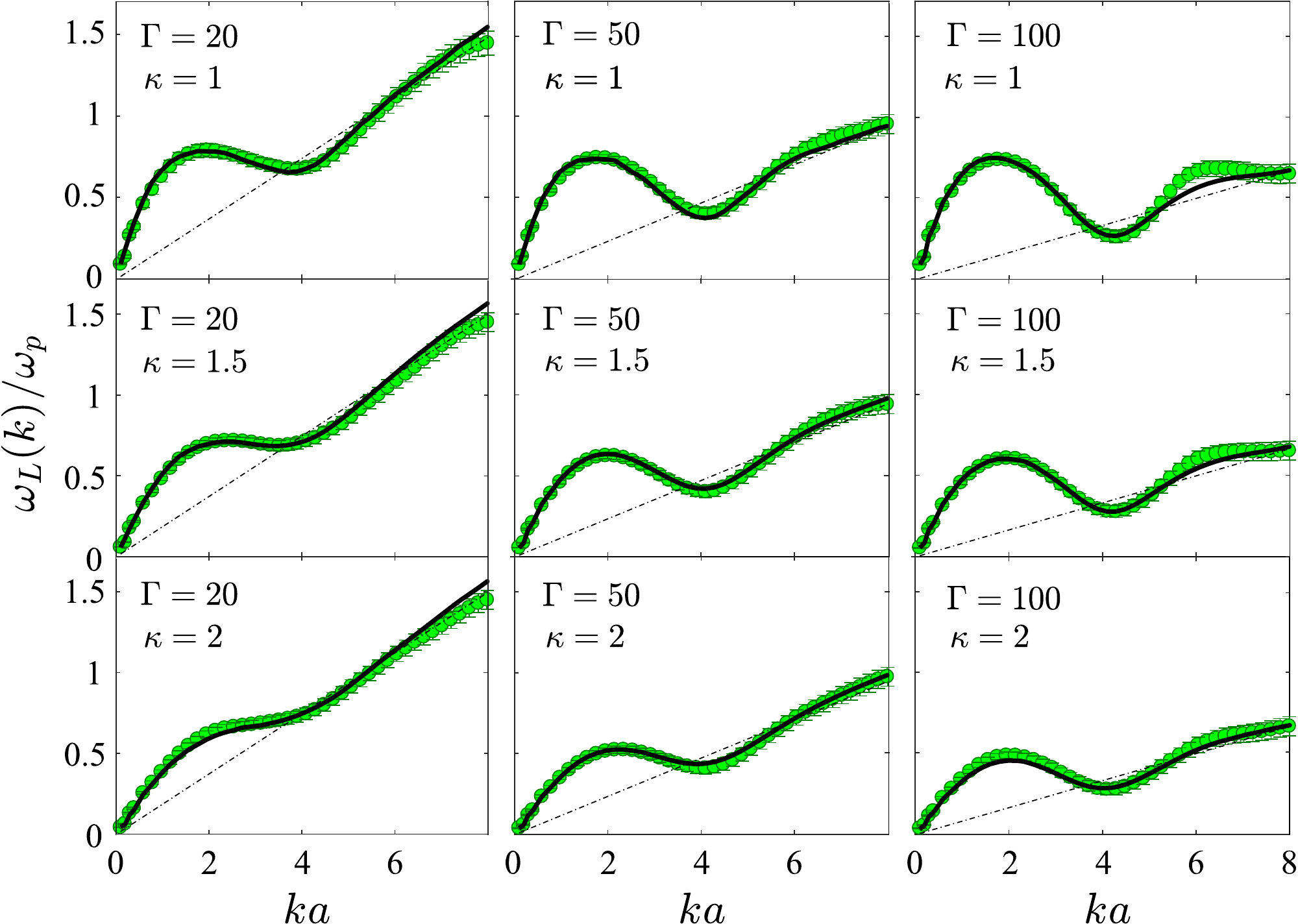} \vskip %
0.3cm
\caption{(\textbf{Top panel}) Dispersions of the Brillouin peak frequency $%
\protect\omega _{c}(k)$ and of the sound dumping coefficient $-\protect%
\delta (k)$ scaled by the plasma frequency $\protect\omega _{p}$; the
results are given for the thermodynamic states ($\Gamma =50,\;\protect\kappa %
=1$) and ($\Gamma =100,\;\protect\kappa =2$). Black solid lines are
theoretical results with Eq.~(\protect\ref{eq: omega_c}) for $\protect\omega %
_{c}(k)$ and with Eq.~(\protect\ref{eq: sound_att}) for $-\protect\delta (k)$%
; red dashed lines depict results of FM-theory~\protect\cite%
{Tkachenko_PRL_2016} and green circles represent MD simulation data.
Dash-dotted lines correspond to the asymptotic results $\protect\omega %
_{c}(k)\sim c_{s}k$ and $-\protect\delta (k)\sim \Gamma _{s}k^{2}$. (\textbf{%
Bottom panel}) Dispersion curves for longitudinal collective excitations
evaluated from Eq.~(\protect\ref{eq: omega_L}) (solid curves) and from MD
simulations (green circles) for nine various thermodynamic ($\Gamma ,\;%
\protect\kappa $)-states. Straight lines correspond to $\protect\omega %
_{L}(k)\sim (a/\protect\sqrt{3\Gamma })k$.}
\label{Fig_dispersions}
\end{figure}
In Fig.~\ref{Fig_dispersions}, we present results characterizing the
capability of the theory to correctly reproduce the high-frequency Yukawa-OCP dynamics. It stems from Fig.~\ref{Fig_dispersions} (top panel) that Eqs.~(%
\ref{eq: omega_c}) and Eq.~(\ref{eq: sound_att}) describe the MD simulations
results for the dispersions characteristics $\omega _{c}(k)$ and $-\delta
(k) $ as well. From the low-$k$ asymptotes $\omega _{c}(k\rightarrow
0)\simeq v_{s}k$ and $\delta (k\rightarrow 0)\simeq -\Gamma _{s}k^{2}$ it is
possible to determine the sound velocity $v_{s}$ and the sound attenuation
coefficient $\Gamma _{s}$. For the conditions pointed out in Fig.~\ref%
{Fig_dispersions} (top panel), we found $v_{s}/(\omega _{p}a)=0.938$ and $%
\Gamma _{s}/(\omega _{p}a^{2})=0.078$ for $\Gamma =50,\;\kappa =1$, and $%
v_{s}/(\omega _{p}a)=0.39$ and $\Gamma _{s}/(\omega _{p}a^{2})=0.083$ for $%
\Gamma =100,\;\kappa =2$. We can conclude that the theory with Eq.~(\ref{eq:
omega_L}) allows one to compute the dispersion curves $\omega _{L}(k)$ for
the longitudinal collective excitations in a wide range of variation of the
system parameters. Full correspondence between the theoretical and the MD
simulation results for the dispersion $\omega _{L}(k)$ is seen in Fig.~\ref%
{Fig_dispersions} (bottom panel), where our results for nine different ($%
\Gamma ,\;\kappa $) combinations are given. The proposed theoretical
description with Eq.~(\ref{eq: omega_L}) properly reproduces the low-$k$ asymptotic
forms and the roton minima located at $ka\simeq 4.3$ under the above
condition. In full agreement with the simulation results, the
theory indicates a smoothing of the roton minimum with a decrease in the
parameter $\Gamma $ and with an increase in $\kappa $, so that the roton
minimum is practically absent when $\Gamma =20$ and $\kappa =2$. The
extremum condition for the dispersion $\omega _{L}(k)$ is
\begin{equation}
\frac{\partial A_{3}(k)}{\partial k}=\omega _{L}^{4}(k)\frac{\partial
A_{1}(k)}{\partial k}.
\end{equation}%
Then, with the known static structure factor $S(k)$ for a specific
combination ($\Gamma ,\;\kappa $), it is possible to accurately predict the
location of the maximum and roton minimum on the dispersion curve $\omega
_{L}(k)$. For example, at $\Gamma =50$ and $\kappa =1.5$, the maximum of $%
\omega _{L}(k)$ is at $ka=1.95$, whereas the minimum is at $ka=4.3$. Note
that the position of the maximum in the dispersion~$\omega _{L}(k)$
approximately coincides with the first pseudo-Brillouin zone boundary $%
k=k_{m}/2$, which corresponds to the transition range from the collective
particle dynamics to the dynamics within an area formed by the neighboring
particles. Moreover, the dispersion law $\omega _{L}(k)$ for all considered
pairs ($\Gamma ,\;\kappa $) demonstrate correct asymptotic forms at large
wavenumbers into the regime of a free-particle dynamics: $\omega _{L}(k)=%
\sqrt{2/(3\Gamma )}\omega _{p}ka$.

In conclusion, it is shown that the self-consistent relaxation theory can be applied to describe the collective dynamics of ions in a strongly coupled classical one-component Yukawa plasma. For the intermediate screening regime of this system, when the interparticle interaction is realized on a finite scale, a correspondence between the sum rules is found, which directly gives analytical expressions for the main characteristics of the collective dynamics, determined through the static structure factor and without any adjustment to the dynamic data.
The present approach generates correct results within the range of parameters, where relations~(\ref{eq: Delta_s}) are satisfied. Precisely, accurate results are produced for the Yukawa-OCP states with the values of the coupling and structure parameters varying in the ranges $\Gamma \in [20;\, 100]$ and $\kappa \in [1;\, 2]$, where the central (Rayleigh) peak is not pronounced. We emphasize that this study does not cover the states of the system corresponding to the weak coupling regime with the coupling parameter $\Gamma \sim 1$ or smaller and to the Coulomb system with  $\kappa \to 0$.

The presented theory is an alternative to the FM-theory presented recently~\cite%
{Tkachenko_PRL_2016}. Both approaches are non-perturbative, and a strong
correspondence between them is elucidated. Energy dissipation processes in
the system under scrutiny are taken into account in these two theoretical
constructions so that they might be considered generalizations of the
quasi-localized charge approximation \cite{Golden}. Conclusions with respect
to the analiticity of the system direct dielectric function envisaged in
\cite{DKM} and elaborated in \cite{MKHD} are confirmed (details are provided
in the Supplemental Material~\cite{Suppl}).

\vskip 0.5 cm

\acknowledgments This work was supported by the Russian Science Foundation
(Project No. 19-12-00022). I.M.T. acknowledges the financial support
provided by the Committee of Science of the Ministry of Education and
Science of the Republic of Kazakhstan (Project No. AP09260349). A.V.M. acknowledges the Foundation for the Advancement of Theoretical Physics and Mathematics ``BASIS'' for supporting the computational part of this work. The authors are grateful to R.M. Khusnutdinoff and B.N. Galimzyanov for discussion of the results of molecular dynamics simulations.

\newpage

\section*{Supplemental Material}

I. Here we wish to clarify the interrelations between the self-consistent
relaxation theory we apply in the main text and the self-consistent version
of the moment approach called there the FM-theory.

Let us start with the fluctuation-dissipation theorem (FDT). If we denote
the inverse dielectric function as $\eta \left( q,\omega \right) $, then the
FDT classical version can be written as (in this Supplemental Material we
use the dimensionless wavenumber $q=ka$, $a$ being the Wigner-Seitz radius)%
\begin{equation*}
\text{Im}\eta \left( q,\omega \right) =-\pi \phi \left( q\right) \beta
\omega S\left( q,\omega \right) \ ,
\end{equation*}%
where $\phi \left( q\right) =4\pi \left( ea\right) ^{2}/q^{2}$ and $\beta $
is the inverse temperature in energy units. As in the main text, we employ
the power frequency moments of the dynamic structure factor (DSF) $S\left(
q,\omega \right) $:%
\begin{equation*}
S_{l}\left( q\right) =\frac{1}{\rho }\int_{-\infty }^{\infty }\omega
^{l}S\left( q,\omega \right) d\omega \ .
\end{equation*}

The FM-theory expression for the DSF \cite{Krein, Adamyan, MB,
TkachenkoPRE2014, Tkachenko_PRL_2016, TkachenkoPRE2020}%
\begin{equation}
\frac{\pi S\left( q,\omega \right) }{\rho S\left( q\right) }=\frac{\omega
_{1}^{2}\left( \omega _{2}^{2}-\omega _{1}^{2}\right) h}{\omega ^{2}\left(
\omega ^{2}-\omega _{2}^{2}\right) ^{2}+h^{2}\left( \omega ^{2}-\omega
_{1}^{2}\right) ^{2}}  \label{1}
\end{equation}%
is a direct consequence of Nevanlinna's theorem as applied to the truncated
Hamburger problem of five moments $\left\{ S_{0},0,S_{2},0,S_{4}\right\} $,
\begin{equation}
\frac{1}{\rho }\int_{.\infty }^{\infty }\frac{S\left( q,\omega ^{\prime
}\right) d\omega ^{\prime }}{\omega -\omega ^{\prime }}=\frac{E_{3}\left(
\omega ;q\right) +Q\left( \omega ;q\right) E_{2}\left( \omega ;q\right) }{%
D_{3}\left( \omega ;q\right) +Q\left( \omega ;q\right) D_{2}\left( \omega
;q\right) }\ .  \label{2}
\end{equation}%
This theorem relates a response function (in mathematics, such functions are
said to belong to the Nevanlinna class $\mathfrak{R}$ \cite{Krein}), which
is the DSF for a classical system, to the Nevanlinna parameter function $%
Q\left( \omega ;q\right) \subset \mathfrak{R}_{0}$ (this is a subclass of $%
\mathfrak{R}$ with an additional condition on the limiting form of the
function:%
\begin{equation}
\lim_{z\rightarrow \infty }Q\left( z;q\right) /z=0  \label{as}
\end{equation}%
to be valid in the upper half-plane of the complex frequency $z=\omega
+i\delta ,\ \delta >0$; this requirement guarantees the fulfillment of the
moment conditions regardless of the specific form of $Q\left( z;q\right) $).
The coefficients of the fractional-linear transformation (\ref{2}), $%
D_{2}\left( \omega ;q\right) =\left( \omega ^{2}-\omega _{1}^{2}\left(
q\right) \right) $ and $D_{3}\left( \omega ;q\right) =\omega \left( \omega
^{2}-\omega _{2}^{2}\left( q\right) \right) $ are orthogonal polynomials
with the even weight $S\left( q,\omega \right) /\rho $:
\begin{equation*}
\frac{1}{\rho }\int_{.\infty }^{\infty }D_{2}\left( \omega ;q\right)
D_{3}\left( \omega ;q\right) S\left( q,\omega \right) d\omega =0\ ,
\end{equation*}%
while the polynomials $E_{3}\left( \omega ;q\right) $ and $E_{2}\left(
\omega ;q\right) $ are their conjugate ones:%
\begin{equation*}
E_{j}\left( \omega ;q\right) =\frac{1}{\rho }\int_{.\infty }^{\infty }\frac{%
D_{2}\left( \omega ^{\prime };q\right) -D_{2}\left( \omega ;q\right) }{%
\omega ^{\prime }-\omega }S\left( q,\omega ^{\prime }\right) d\omega
^{\prime }\ ,\quad j=1,2\ ,
\end{equation*}%
that is
\begin{equation*}
E_{3}\left( \omega ;q\right) =S_{0}\left( q\right) \left( \omega
_{1}^{2}\left( q\right) -\omega _{2}^{2}\left( q\right) +\omega ^{2}\right)
\end{equation*}%
and
\begin{equation*}
E_{2}\left( \omega ;q\right) =\omega S_{0}\left( q\right) \ .
\end{equation*}

It should be emphasized that the theorem (\ref{2}) gives the most accurate
expression for the DSF within the five-moment Hamburger problem: restore the
DSF from the moments $\left\{ S_{0},0,S_{2},0,S_{4}\right\} $, where $%
S_{0}\left( q\right) =S\left( q\right) $ is the system static structure
factor. The explicit expressions for the moments are well-known, in
particular, it can be shown that the expression for the second-order
relaxation parameter
\begin{equation}
\Delta_{2}(q)=\Delta_{1}(q)\left[ 3S(q)-1\right] +\omega_{p}^{2}\,D(q)
\label{Delta_2}
\end{equation}
can be rewritten as
\begin{equation}
\Delta_{2}(q)=\omega_{p}^{2}\left[ \frac{q^{2}}{q^{2}+\kappa^{2}}+\frac{q^{2}%
}{\Gamma}-\frac{q^{2}}{3\Gamma S\left( q\right) }+U\left( q\right) \right] \
,  \label{Delta21}
\end{equation}
where two equivalent forms of the interaction contribution $U\left( q\right)
$ can be employed \cite{Balucani}
\begin{equation*}
U\left( q\right) =\int_{0}^{\infty}\frac{\exp\left( -\kappa x\right) }{3x} \left( g\left( x\right)-1\right)f_1\left(q,x,\kappa\right) dx
\end{equation*}
with
\begin{align*}
f_1\left(q,x,\kappa\right)= \left( 2\left( \kappa x\right) ^{2}+6\kappa x+6\right) j_{2}\left(qx\right) +\\
+\left( \kappa x\right) ^{2}\left( 1-j_{0}\left( qx\right) \right) 
\end{align*}
or \cite{Tkachenko_PRL_2016}%
\begin{equation*}
U\left( q\right) =\frac{1}{12\pi}\int_{0}^{\infty}\left[ S\left( p\right) -1%
\right] f_2\left( p,q,\kappa\right) p^{2}dp\ ,
\end{equation*}
with
\begin{align*}
f_2\left( p,q,\kappa\right) =\frac{2\left( 3q^{2}-\kappa^{2}-p^{2}\right) }{%
q^{2}}+\frac{\left( q^{2}-\kappa^{2}-p^{2}\right) ^{2}}{2q^{3}p}\times\\\times\ln\left(
\frac{\kappa^{2}+\left( q+p\right) ^{2}}{\kappa^{2}+\left( q-p\right) ^{2}}%
\right) -\frac{8p^{2}}{3\left( \kappa^{2}+p^{2}\right) }\ .
\end{align*}
Here $h(x)=g(x)-1$, $g(x)$ is the radial distribution function. The
frequency relaxation parameters $\Delta_1(q)$ and $\Delta_2(q)$ are related
to the characteristic frequencies $\omega_1^2(k)$ and $\omega_2^2(k)$ by the
following relations:
\begin{eqnarray}
\omega _{1}^{2}(q) &=& \Delta _{1}(q), \\
\omega _{2}^{2}(q) &=& \Delta _{1}(q) + \Delta _{2}(q) .
\end{eqnarray}

The Kramers-Kronig relations for the inverse dielectric function $\eta\left(
q,z\right) =\epsilon^{-1}\left( q,z\right) $ and the FDT should be applied
to make use of the exact mathematical result (\ref{2}):%
\begin{align*}
\eta\left( q,\omega+i0^{+}\right) & =1+\frac{1}{\pi}\int_{.\infty}^{\infty}%
\frac{\text{Im}\eta\left( q,\omega^{\prime}\right) d\omega^{\prime}}{%
\omega^{\prime}-\omega-i0^{+}}= \\
& =1-\phi\beta\rho\left( S_{0}\left( q\right) -\frac{\omega}{\rho}%
\int_{.\infty}^{\infty}\frac{S\left( q,\omega^{\prime}\right) d\omega
^{\prime}}{\omega-\omega^{\prime}}\right) = \\
& =1+\frac{\phi\beta\rho S\left( q\right) \omega_{1}^{2}\left( q\right)
\left( \omega+Q\left( \omega;q\right) \right) }{\omega\left( \omega
^{2}-\omega_{2}^{2}\left( q\right) \right) +Q\left( \omega;q\right) \left(
\omega^{2}-\omega_{1}^{2}\left( q\right) \right) }\ .
\end{align*}
Thus,
\begin{equation}
S\left( q,\omega\right) =\frac{\rho S\left( q\right) }{%
\pi\omega}\text{Im}\frac{-\omega_{1}^{2}\left( \omega+Q\left( \omega;q\right) \right) }{%
\omega\left( \omega^{2}-\omega_{2}^{2}\left( q\right) \right) +Q \left( \omega^{2}-\omega_{1}^{2}\left( q\right) \right) }\ .
\label{3}
\end{equation}
Notice that the latter expression can be rewritten as a truncated continuous
fraction:%
\begin{equation}
\frac{\pi S\left( q,\omega\right) }{\rho S\left( q\right) }=\frac {1}{\omega}%
\text{Im}\left( 1-\frac{\omega}{\omega-\frac{\omega _{1}^{2}}{\omega-\frac{%
\omega_{2}^{2}-\omega_{1}^{2}}{\omega+Q}}}\right) \ .  \label{cp}
\end{equation}

In \cite{Adamyan, MB, TkachenkoPRE2014, Tkachenko_PRL_2016, TkachenkoPRE2020}
the Nevanlinna parameter function $Q\left( z;q\right) $ was approximated by
its static value $Q\left( z;q\right) =Q\left( 0^{+};q\right) =ih\left(
q\right) $. In this case%
\begin{align}
\frac{\pi S\left( q,\omega \right) }{\rho S\left( q\right) }& =-\frac{\omega
_{1}^{2}}{\omega }\text{Im}\frac{\left( \omega +ih\right) }{\omega \left(
\omega ^{2}-\omega _{2}^{2}\right) +ih\left( \omega ^{2}-\omega
_{1}^{2}\right) }=  \notag \\
& =\frac{\omega _{1}^{2}h\left( \omega _{2}^{2}-\omega _{1}^{2}\right) }{%
\omega ^{2}\left( \omega ^{2}-\omega _{2}^{2}\right) ^{2}+h^{2}\left( \omega
^{2}-\omega _{1}^{2}\right) ^{2}}\ .  \label{4}
\end{align}

In \cite{Tkachenko_PRL_2016} on the basis of some physical considerations
the positive static parameter function $h\left( q\right) $ was related to
the characteristic frequencies $\omega_{1}\left( q,\kappa\right) $ and $%
\omega_{2}\left( q,\kappa\right) $, which permitted to express the dynamic
properties of one-component classical plasmas to their static
characteristics, i.e., the static structure factor.

In the present work the function $Q\left( z;q\right) $ is obtained within
the Zwanzig-Mori formalism as \cite{Mokshin/Galimzyanov_JPCM_2018}
\begin{equation}
\tilde{Q}\left( z;q\right) =-\frac{\Delta _{3}(q)}{2\Delta _{4}(q)}\left( z-%
\sqrt{z^{2}-4\Delta _{4}(q)}\right) \ .  \label{Q1}
\end{equation}%
This model Nevanlinna function satisfies the asymptotic requirement (\ref{as}%
), but it is obviously not analytical in the upper half-plane of the
variable $z$, i.e., $\tilde{Q}\left( z;q\right) \notin $ $\mathfrak{R}$.
Nevertheless, if we substitute into (\ref{cp})
\begin{equation}
\tilde{Q}\left( \omega ;q\right) =-\frac{\Delta _{3}(q)}{2\Delta _{4}(q)}%
\left( \omega -i\sqrt{4\Delta _{4}(q)-\omega ^{2}}\right) \ ,  \label{Q2}
\end{equation}%
for the frequencies such that $\omega ^{2}\ll \Delta _{4}(q)$, it stems from
(\ref{cp}) and (\ref{Q2}) that
\begin{equation}
\frac{\pi S\left( q,\omega \right) }{\rho S\left( q\right) }=\frac{\Delta
_{1}(q)\Delta _{2}(q)\Delta _{3}(q)
\sqrt{\Delta _{4}(q)}/\left(\Delta _{4}(q)-\Delta _{3}(q)\right)}{\omega ^{6}+\mathcal{A}_{1}(q)\omega ^{4}+\mathcal{A}%
_{2}(q)\omega ^{2}+\mathcal{A}_{3}(q)}\ .  \label{SZ}
\end{equation}%
with
\begin{equation*}
\mathcal{A}_{1}(q)=\frac{\Delta _{3}^{2}(q)-\Delta _{2}(q)\left[ 2\Delta
_{4}(q)-\Delta _{3}(q)\right] }{\Delta _{4}(q)-\Delta _{3}(q)}-2\Delta
_{1}(q)\ ,
\end{equation*}%
\begin{align*}
\mathcal{A}_{2}(q)=\frac{\Delta _{2}^{2}(q)\Delta _{4}(q)-2\Delta
_{1}(q)\Delta _{3}^{2}(q) }{\Delta _{4}(q)-\Delta _{3}(q)}+\Delta
_{1}^{2}(q)+\\
+\frac{\Delta _{1}(q)\Delta _{2}(q)\left[ 2\Delta
_{4}(q)-\Delta _{3}(q)\right]}{\Delta _{4}(q)-\Delta _{3}(q)}\ ,
\end{align*}%
\begin{equation*}
\mathcal{A}_{3}(q)=\frac{\Delta _{1}^{2}(q)\Delta _{3}^{2}(q)}{\Delta
_{4}(q)-\Delta _{3}(q)}\ .
\end{equation*}%
Observe that the coefficients $\mathcal{A}_{j}(q)$ coincide with the
coefficients $A_{j}(q)$, $j=1,2,3$ introduced in the main text. Notice also
that when the conditions
\begin{equation*}
\omega ^{2}\ll \Delta _{4}(q)
\end{equation*}%
and
\begin{equation*}
\left[ \Delta _{1}\left( q\right) +\Delta _{2}\left( q\right) \right]
^{2}\Delta _{4}\left( q\right) =2\Delta _{1}\left( q\right) \Delta
_{3}^{2}\left( q\right)
\end{equation*}%
are satisfied, expressions for the DSF, (\ref{SZ}) and (\ref{4}) coincide
and when $\Delta _{3}\left( q\right) =0,$ they possess joint poles
\begin{equation*}
z_{\pm }=\pm \omega _{2}\left( q\right) =\pm \sqrt{\Delta _{1}\left(
q\right) +\Delta _{2}\left( q\right) }\ .
\end{equation*}%
It is also very important that the asymptotic forms of the inverse
dielectric function related to these two expressions for the DSF also
coincide, i.e. they satisfy the involved sum rules automatically,
independently of the chosen model expression for the Nevanlinna function.

The overall quality of both outlined approaches is corroborated by the
comparison with the simulation data carried out in the main text.

II. The self-consistent relaxation theory allows one to obtain the
dispersion equation in the analytical form (see equation (11) in the main
text). Then it turns out to be appropriate to discuss the issue of
analyticity of the direct dielectric function like it was done in the works
\cite{DKM} and \cite{MKHD}. If we make the substitution $s=-\text{i}z$,
equation (11) in the main text takes the following form:
\begin{equation}
z\biggl(z^{2}-\Delta _{1}(q)-\frac{8}{5}\Delta _{2}(q)\biggr)+\text{i}%
B(q)(z^{2}-\Delta _{1}(q))=0,  \label{inv_df_dip}
\end{equation}%
whose roots $z=\pm \omega _{c}-\text{i}\delta $ are the poles of the inverse
dielectric function of the system, $\eta (q,z)$ \cite{TkachenkoPRE2020}. The
location of these poles in the lower half-plane of the complex frequency
implies that the sound attenuation coefficient in the system under
investigation is positive and thus it is stable \cite{DKM}. The last
condition is fulfilled for all thermodynamic states considered in this work.
If in the equation (\ref{inv_df_dip}) we make the replacement $\Delta _{1}(q)
$ = $\Delta _{1}(q)-\omega _{p}^{2}$, then roots with a positive imaginary
part will appear in its solution, which corresponds to a negative sound
attenuation coefficient and the system instability. This behavior of
solutions to the equation (\ref{inv_df_dip}) with this replacement means
that the following inequality holds:
\begin{equation}
\frac{\Delta _{1}(q)}{\omega _{p}^{2}}<1.
\end{equation}%
This condition is equivalent to the negativity of the direct static
dielectric function $\epsilon (q,z=0)$ and means the fulfillment of the
causality principle described in the work \cite{DKM} and studied in detail
for the Coulomb one-component plasma in \cite{MKHD}.

\end{document}